\documentstyle[letter,preprint,epsf,wrapft]{ptptex}

\def\lsim{\ \lower-0.4ex\hbox{$<$}\kern-0.80em\lower0.8ex\hbox{$\sim$}\ }
\def\gsim{\ \lower-0.4ex\hbox{$>$}\kern-0.80em\lower0.8ex\hbox{$\sim$}\ }

\preprintnumber{
UTAP-256\\ RESCEU-19/97\\}
\title{
Quantum Nucleation of Two-Flavor Quark Matter in Neutron Stars}

\author{
Kei {\sc Iida}$^{(a)}$
\footnote{E-mail address: 
iida@utaphp1.phys.s.u-tokyo.ac.jp} 
and Katsuhiko {\sc Sato}$^{(a,b)}$
\footnote{E-mail address: 
sato@phys.s.u-tokyo.ac.jp}
}

\inst{
$^{(a)}$Department of Physics, University of Tokyo, 7-3-1 Hongo, Bunkyo, 
Tokyo 113
\\
$^{(b)}$Research Center for the Early Universe,
University of Tokyo, 7-3-1 Hongo, Bunkyo, Tokyo 113
}

\recdate{
\today
}

\abst{
Rates for nucleation of two-flavor quark matter in a neutron star
core, originally composed of nuclear matter in $\beta$ equilibrium,
are calculated at zero temperature by a quantum tunneling
analysis incorporating the electrostatic energy. 
We find that a nucleated droplet would develop into bulk matter
due to electron screening effects.
}

\begin{document}

\maketitle

The possibility that quark matter could exist in neutron stars and the
astrophysical consequences have been considered for the past two
decades. (See, e.g., Refs.\ 1)-3).) Usually, the main question was
at what pressure the free energy per baryon for electrically neutral
quark matter becomes less than that for nuclear matter.
Recently, Glendenning$^{1)}$ discovered that, by relaxing the constraint 
of $local$ charge neutrality, a phase where quark and nuclear matter 
in $\beta$ equilibrium coexist in a uniform sea of electrons could
appear for a finite range of pressures. This is because the presence of
strange and down quarks plays a role in reducing the electron Fermi
energy and in increasing the proton fraction of nuclear matter. 
His work$^{1)}$ and the
subsequent work by Heiselberg et al.$^{2)}$ claimed that such a
mixed phase could exhibit spatial structure such as quark matter
droplets embedded in nuclear matter due to surface and Coulomb effects. 
As the star whose core consists of nuclear matter in $\beta$ 
equilibrium spins down or accretes matter from its companion star,
the central density could become sufficiently large for the mixed phase to 
be stable. Whether the mixed phase actually nucleates, however, depends on 
the occurrence of the dynamical processes leading to its nucleation. 
The first of these processes should be the conversion of nuclear matter 
to two-flavor quark matter, since the conversion to three-flavor 
quark matter, which requires many simultaneous weak interactions, 
is unlikely to occur.$^{4)}$ In this paper, therefore, 
we consider at what pressure a droplet of two-flavor quark matter 
forms in nuclear matter and whether it develops into bulk matter 
or remains finite. 

We begin by describing the bulk properties of the various components
for densities well above the nuclear saturation density. 
For nucleons,
we write the simple formula for the energy density
adopted by Heiselberg et al.:$^{2)}$ 
\begin{equation}
  \epsilon_{N}(n,x)=n\left[m_{n}+\frac{K_{0}}{18}
  \left(\frac{n}{n_{0}}-1\right)^{2}+S_{0}\left(\frac{n}{n_{0}}
  \right)^{\gamma}(1-2x)^{2}\right]\ ,
\end{equation}
where $m_{n}$ is the neutron mass, $n$ is the nucleon number density, 
$n_{0}=0.16$ fm$^{-3}$ is the nuclear saturation density, 
$x$ is
the proton fraction, $K_{0}=250$ MeV is the coefficient of the compressional 
term, and $S_{0}=30$ MeV as well as $\gamma=1$ determine
the symmetry term.
For quarks we adopt the energy density based on the bag model,
\begin{equation}
  \epsilon_{Q}(n_{u},n_{d},n_{s})=\left(1-\frac{2\alpha_{s}}{\pi}\right)
  \left(\sum_{q=u,d,s}\frac{3\mu_{q}^{4}}{4\pi^{2}}\right)+B\ ,
\end{equation}
where $q=u, d,$ and $s$ denote up, down, and strange quarks, 
$n_{q}$ is the number density of $q$ quarks, and
$\mu_{q}=(1-2\alpha_{s}/\pi)^{-1/3}(\pi^{2}n_{q})^{1/3}$ is the chemical
potential of $q$ quarks. 
In Eq.\ (2), all quark masses have been taken to
be zero, and the QCD fine structure constant and the bag constant
have been set as $\alpha_{s}=0.4$ and $B=120$ MeV fm$^{-3}$. 
The energy density of the electrons, which are relativistically degenerate,
is $\epsilon_{e}(n_{e})=\mu_{e}^{4}/4\pi^{2}$ with the electron 
number density $n_{e}$ and 
the electron 
chemical potential $\mu_{e}=(3\pi^{2}n_{e})^{1/3}$.
We neglect the existence
of hyperons and muons.

\begin{wrapfigure}{r}{6.6cm}   
                \vspace{-0.6cm}
	        \epsfxsize=7.3cm 
	        \epsfysize=8.4cm    
	        \centerline{\epsfbox{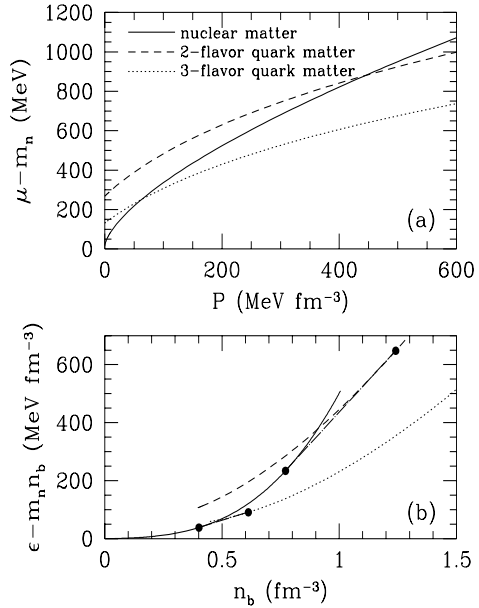}}
          \caption{(a) Chemical potentials for the electrically neutral
bulk phases as a function of pressure. The solid line stands for 
the phase of nuclear matter in $\beta$ equilibrium; the dashed line,  
the phase of $u$ and $d$ quark matter arising therefrom via 
deconfinement; the dotted line, the phase of $u$, $d$, and $s$ quark 
matter in $\beta$ equilibrium. (b) Energy densities for the
corresponding phases as a function of baryon density. The
double-tangent constructions (dot-dash lines) denote the 
coexistence of the two bulk phases involved.}
          \label{fig:1}
\end{wrapfigure}
At a given pressure $P$, we have calculated the baryon chemical 
potentials
for the three bulk phases being electrically neutral.
The first phase consists of nuclear matter in $\beta$
equilibrium: $\mu_{n}=\mu_{p}+\mu_{e}$ with the neutron (proton)
chemical potential $\mu_{n}$ $(\mu_{p})$. The second phase is composed 
of $u$ and $d$ quark matter arising via deconfinement
from the above nuclear matter of proton
fraction $x_{\rm eq}$ so as to satisfy $n_{u}/n_{d}=(1+x_{\rm eq})
/(2-x_{\rm eq})$. Lastly, for comparison,
we consider a phase containing 
$u$, $d$, and $s$ quark matter in $\beta$ equilibrium: 
$\mu_{d}=\mu_{u}+\mu_{e}$ and $\mu_{d}=\mu_{s}$. 
Figure 1(a) shows the chemical potentials for these three phases
obtained as 
$\mu=(\epsilon+P)/n_{b}$ with the corresponding total energy densities 
$\epsilon$ and baryon densities $n_{b}$. The crossings denote
the transition points from one bulk phase to another, at which 
the baryon density jumps as can be seen from Fig.\ 1(b).

Let us now inquire how long it takes a $u$ and $d$ quark matter droplet
to form via deconfinement in $\beta$-equilibrium nuclear matter 
for pressures in the vicinity of the transition point shown in Fig.\ 1(a).
We assume such a droplet to be a sphere macroscopically characterized
by its radius $R$ and describe 
the tunneling behavior of a virtual droplet 
in the semiclassical approximation. A Lagrangian for the
fluctuating droplet may thus be written as$^{5)}$
\begin{equation}
   L=\frac{1}{2}M(R){\dot R}^{2}-U(R)\ ,
\end{equation}
where $M(R)$ is the effective droplet mass, and $U(R)$ is the
potential for droplet formation. In the present analysis, no dissipation
is considered.   

We proceed to evaluate the potential $U(R)$. Due to the high sound 
velocity of the system ($\sim c$) 
it may be assumed that the number density of each 
component adjusts adiabatically to fluctuations of $R$. Consequently, 
the system retains pressure equilibrium between quark and nuclear matter,
and each component, if it makes no contribution to Coulomb screening on
the resulting excess droplet charge, is homogeneous, or else
is distributed accordingly. 
Here we assume the quark distributions to be uniform; their shifts
due to quark screening$^{6)}$ should
reduce the Coulomb energy only slightly.
The electron gas is most effective 
at screening; its Thomas-Fermi (TF) screening length 
is $\sim 6$ fm since $\mu_{e}\sim 300$ MeV. 
For the droplet sizes of interest 
$R\lsim 10$ fm, protons, whose TF screening length is estimated from
Eq.\ (1) as $\gsim 10$ fm, 
may be taken to be uniform. For the initial metastable phase 
of nuclear matter in $\beta$ equilibrium
under pressure $P$, we set the chemical potential 
as $\mu_{\rm init}=(\epsilon_{\rm init}+P)/n_{\rm init}$ with the energy 
density $\epsilon_{\rm init}=\epsilon_{N}(n_{\rm init},x_{\rm init})
+\epsilon_{e}(n_{e,{\rm init}})$, where $n_{\rm init}$, $x_{\rm init}$,
and $n_{e,{\rm init}}$ give the initial number densities of nucleons
and electrons. 
We can thus express the energy density 
for the inhomogeneous phase containing a single droplet
as the sum of bulk, surface, and Coulomb terms: 
\begin{equation}
  \epsilon_{D}=\theta(R-r)\epsilon_{Q}(n_{u},n_{d},0)
   +\theta(r-R)\epsilon_{N}(n_{\rm init},x_{\rm init}) 
   +\epsilon_{e}[n_{e}(r)]+\epsilon_{S}(r)
   +\frac{E(r)^{2}}{8\pi}\ ,
\end{equation}
where 
$r$ is the distance from the center of the droplet,  
$\epsilon_{S}(r)$ is the increase in energy density due to the
quark and nucleon distributions in the surface layer
whose thickness we assume to be much smaller than $R$,
and $E(r)$ is the electric field. 
We then integrate the difference in the thermodynamic potential per
unit volume at chemical potential $\mu_{\rm init}$
between the initial and inhomogeneous phases over the system volume $V$
which is related to $P$ as $P=-\partial(V\epsilon_{\rm init})/\partial V$.
We thus obtain $U(R)$ as
\begin{equation}
   U(R)=\frac{4\pi R^{3}}{3}n_{b,Q}(\mu_{Q}-\mu_{\rm init})
   +\Delta E_{e}(R)+4\pi\sigma R^{2}+E_{C}(R)\ ,
\end{equation}
where $n_{b,Q}=(n_{u}+n_{d})/3$ is the baryon density inside the
droplet, $\mu_{Q}=[\epsilon_{Q}(n_{u},n_{d})+\epsilon_{e}(n_{e,{\rm init}})
+P]/n_{b,Q}$ is the chemical potential for the uniform quark matter 
determined by $n_{u}$, $n_{d}$, $n_{e,{\rm init}}$, and $P$,
$\Delta E_{e}(R)=\int_{V}dV\{\epsilon_{e}[n_{e}(r)]
-\epsilon_{e}(n_{e,{\rm init}})\}$ is the excess of electron energy 
over the initial one, $\sigma$ is the surface tension,
and $E_{C}(R)$ is the electrostatic energy. 
In Eq.\ (5) the curvature energy $(\propto R)$ 
is omitted.
Estimates of the quantities $n_{u}$, $n_{d}$, $n_{e}(r)$, 
$\Delta E_{e}(R)$, $\sigma$, and $E_{C}(R)$ are as follows.

The distribution of electrons has been determined in the linear TF
approximation.$^{7)}$ Electrons act to screen the excess droplet charge
$Ze=(4\pi R^{3}/3)\rho_{Q}$ with $\rho_{Q}=e[(2n_{u}-n_{d})/3
-n_{e,{\rm init}}]$, and thereby deviate from uniformity 
in proportion to the electrostatic potential $\phi(r)$ as
$\delta n_{e}(r)=n_{e}(r)-n_{e,{\rm init}}=
(\partial n_{e}/\partial\mu_{e})_{\rm init}e\phi(r)$.
The potential $\phi(r)$ satisfies the Poisson equation
$\nabla^{2}\phi(r)-\kappa^{2}\phi(r)=-4\pi\rho_{Q}\theta(R-r)$,
where 
$\kappa=\sqrt{4\pi e^{2}(\partial n_{e}/\partial\mu_{e})_{\rm init}}$
is the inverse of the TF screening length. 
Its 
solution 
reads as
\begin{equation}
  \phi(r)=\left\{ \begin{array}{lll}
        (4\pi\rho_{Q}/\kappa^{2})
        [1-\exp(-\kappa R)(1+\kappa R)\sinh(\kappa r)/\kappa r]\ ,
         & \mbox{$r\leq R$} \\
             \\
        (4\pi\rho_{Q}/\kappa^{2})[\kappa R\cosh(\kappa R)
         -\sinh(\kappa R)]\exp(-\kappa r)/\kappa r\ ,
         & \mbox{$r>R\ .$}
 \end{array} \right.
\end{equation}
Figure 2 exhibits a typical example of the electron screening at
$R\sim 3$ fm. When the initial chemical potential of electrons
$\mu_{e,{\rm init}} \gg e\phi$, as in Fig.\ 2, 
the linear TF approximation can be safely used. 
For $R\lsim 10$ fm, it turns out that
$e\phi/\mu_{e,{\rm init}}\lsim 0.1$. 
Within the confines of this approximation,
$\Delta E_{e}(R)=Z\mu_{e,{\rm init}}$.

We then calculate the electrostatic energy $E_{C}(R)$ 
using Eq.\ (6). The result is
\begin{equation}
  E_{C}(R)=(2\pi^{2}\rho_{Q}^{2}/\kappa^{5})
  \{-3+(\kappa R)^{2}+\exp(-2\kappa R)[3+6\kappa R+5(\kappa R)^{2}
   +2(\kappa R)^{3}]\}\ .
\end{equation}
Note that, in the limit of electron uniformity
($\kappa R\rightarrow 0$), $E_{C}(R)\rightarrow 3Z^{2}e^{2}/5R$. 

\begin{wrapfigure}{r}{6.6cm}   
                \vspace{-0.5cm}
	        \epsfxsize=7.3cm 
	        \epsfysize=6.3cm   
	        \centerline{\epsfbox{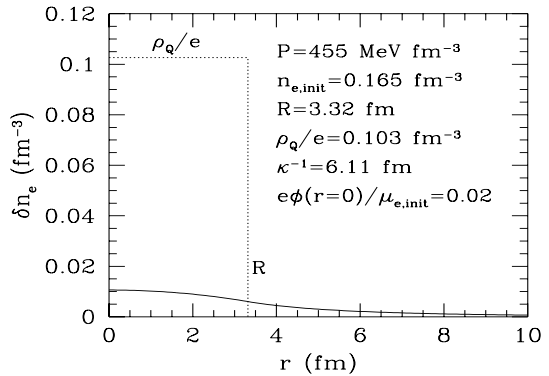}}
       \caption{Deviation $\delta n_{e}$ of the electron number density 
from the uniform one $n_{e,{\rm init}}$ due to the TF screening 
on the excess droplet charge of density $\rho_{Q}$ and of radius 
$R$ at $P=455$ MeV fm$^{-3}$. See the text for the definition of
$r$, $\kappa$, $\mu_{e,{\rm init}}$, and $\phi$.}
          \label{fig:2}
\end{wrapfigure}
The surface tension $\sigma$, being poorly known, has been taken
from the Fermi-gas model
for a quark matter droplet in vacuum.$^{8)}$ Due to the reduction
in quark density of states arising from the droplet surface,
the surface tension $\sigma=(3/4\pi^{2})(m_{u}\mu_{u}^{2}+
m_{d}\mu_{d}^{2})$ occurs for $\mu_{q}\gg m_{q}$ ($q=u, d$). 
Since $\mu_{u}\sim 500$ MeV, $\mu_{d}\sim 600$ MeV, and
$m_{u}\sim m_{d}\sim 10$ MeV, the value of $\sigma$ is
roughly estimated as $\sigma\sim 10$ MeV fm$^{-2}$. 
The Fermi-gas description of finite quark matter as adopted here, 
however, is problematic for a small baryon number inside the bag 
($\lsim 5$).$^{9)}$ Nevertheless, we expect such a description to be 
valid, since the baryon number of a virtual 
droplet moving under the potential barrier is 
$\gsim 10$ (typically $\sim100$).

Finally, we determine the number densities $n_{u}$ and $n_{d}$. 
As a result of deconfinement, these are related via 
$n_{u}/n_{d}=(1+x_{\rm init})/(2-x_{\rm init})$.  
Another relation comes from the pressure equilibrium 
between quark and nuclear matter,
\begin{equation}
   P_{Q}=P_{N}-\frac{\Delta E_{e}(R)}{4\pi R^{3}}\ ,
\end{equation}
where $P_{Q}=[\epsilon_{Q}(n_{u},n_{d})-4B]/3$ is the quark pressure, 
and $P_{N}=P-\epsilon_{e}(n_{e,{\rm init}})/3$ is the nucleon
pressure. As long as $\sigma\propto n_{q}^{2/3}$, the surface tension 
contributes nothing to the pressure equilibrium (8). In Eq.\ (8)
the Coulomb pressure arising from $E_{C}(R)$ has been ignored since
it is negligibly small. $n_{u}$ and $n_{d}$ are thus 
independent of $R$. 

We next estimate the effective mass $M(R)$ from the kinetic energies
of nucleons and of electrons. We obtain the nucleon kinetic energy 
by solving the nucleon continuity equation 
with the boundary condition at the 
droplet surface$^{5)}$ as $2\pi\epsilon_{N}(n_{\rm init},x_{\rm init})
(1-n_{b,Q}/n_{\rm init})^{2}R^{3}{\dot R}^{2}$. The electron kinetic
energy, which is calculated from the electron continuity equation
as $<2\pi\epsilon_{e}(n_{e,{\rm init}})
(\rho_{Q}/n_{e,{\rm init}}e)^{2}R^{3}{\dot R}^{2}$, proves
negligible since $\epsilon_{N}
\sim 1000$ MeV fm$^{-3}$ $\gg\epsilon_{e}
\sim 40$ MeV fm$^{-3}$.  
We thus obtain
\begin{equation}
   M(R)=4\pi\epsilon_{N}(n_{\rm init},x_{\rm init})
\left(1-\frac{n_{b,Q}}{n_{\rm init}}\right)^{2}R^{3}\ .
\end{equation}

\begin{figure}[b]
\parbox{\halftext}{
	        \epsfxsize=7.3cm 
	        \epsfysize=6.3cm 
	        \centerline{\epsfbox{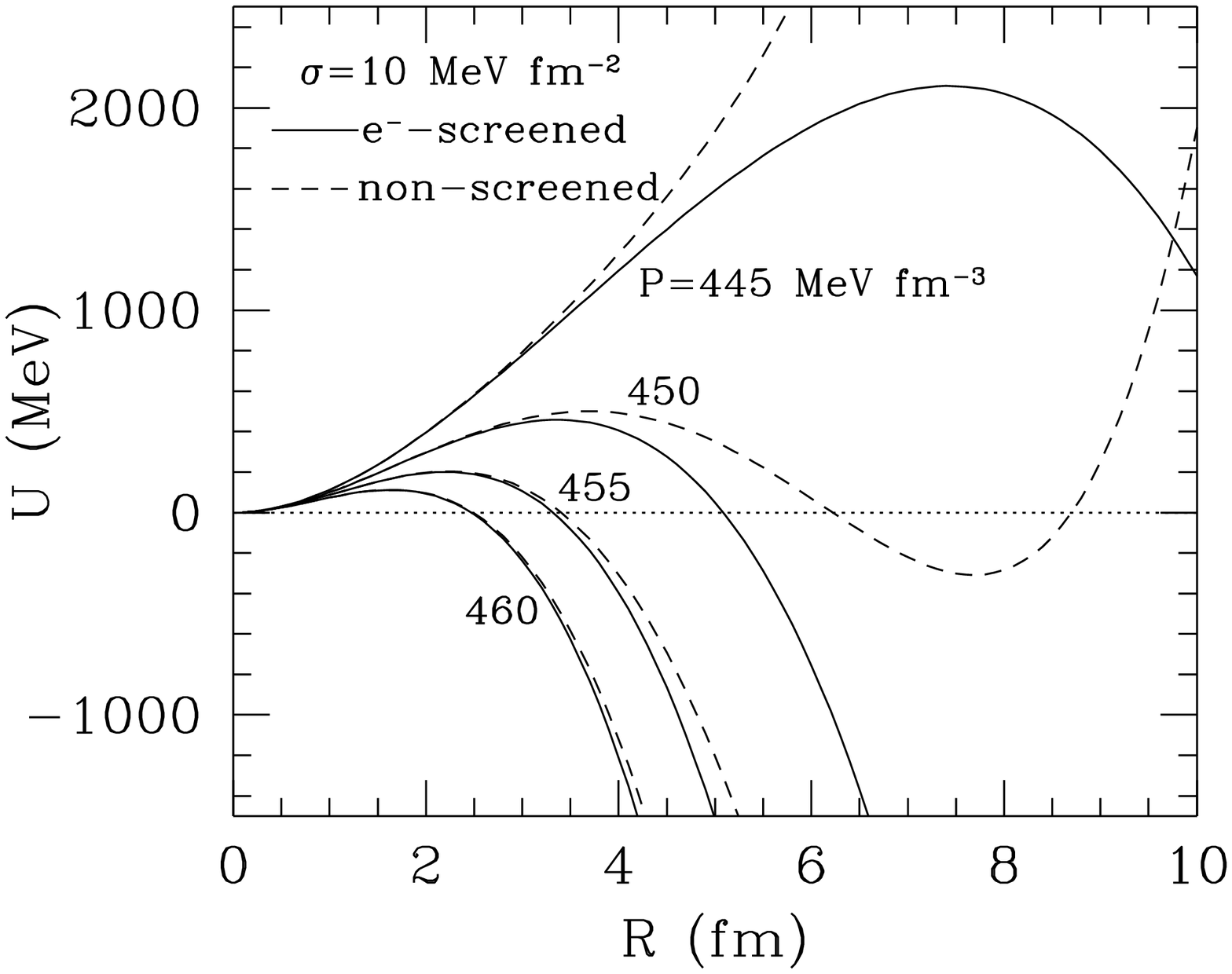}}
                \caption{
Potentials for the formation of a droplet with the surface 
tension $\sigma=10$ MeV fm$^{-2}$ at $P=445, 450, 455$, and $460$ MeV 
fm$^{-3}$. The solid lines are the results for the electron-screened 
droplet and the dashed lines are those for the non-screened droplet.}}
          \label{fig:3}
\hspace{8mm}
\parbox{\halftext}{
	        \epsfxsize=7.3cm 
	        \epsfysize=6.3cm 
	        \centerline{\epsfbox{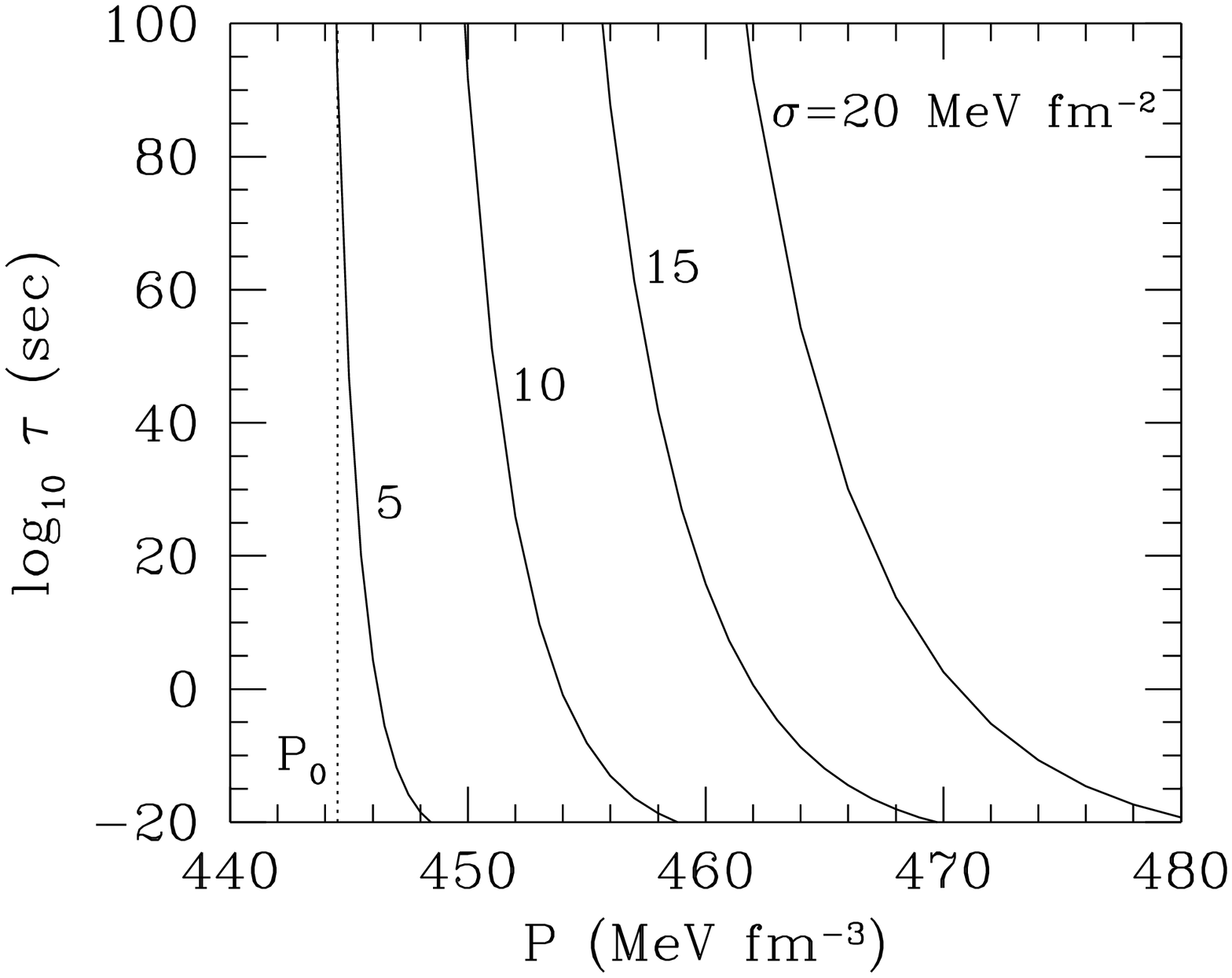}}
       \caption{Time for the formation of a single droplet with the
surface tension $\sigma=5, 10, 15,$ and $20$ MeV fm$^{-2}$ as a
function of pressure. 
$P_{0}$ denotes the transition point.}}
          \label{fig:4}
\end{figure}
We turn to the evaluations of 
the time $\tau$ required to form a single droplet.
By calculating not only the energy $E_{0}$ for the zeroth bound state 
around $R=0$ but also the corresponding oscillation frequency $\nu_{0}$ 
and probability of barrier penetration $p_{0}$
from the Lagrangian (3) in the WKB approximation,$^{10)}$
the values of $\tau$ have been obtained 
as $\tau=(\nu_{0}p_{0})^{-1}$. With increasing pressure
from the transition point $P_{0}\cong445$ MeV fm$^{-3}$, the potential
barrier is lowered for fixed $\sigma$
as the solid lines in Fig.\ 3 exhibit. 
Since this barrier controls the underbarrier action $S_{0}$ between
the classical turning points and hence the probability 
$p_{0}=\exp(-2S_{0}/\hbar)$,
the nucleation time $\tau$ calculated for
$\sigma=5, 10, 15,$ and $20$ MeV fm$^{-2}$ shows an exponential $P$ 
dependence in the overpressure regime, as illustrated in Fig.\ 4. 
The oscillation time scale is naturally $\nu_{0}^{-1}\sim10^{-23}$ sec.
The energy $E_{0}$ is mostly comparable to the barrier height 
$\sim 100$ MeV. 
Nevertheless, the WKB approximation is expected to be useful because
the WKB analysis of tunneling for the 
Lagrangian having no Coulomb energy but otherwise the same
$R$ dependence as Eq.\ (3)
agrees well with the fully quantum analysis.$^{11)}$
It is important to note that the formation time of the first droplet 
anywhere in a neutron star is $\tau/N$, where $N$
is the number of virtual centers of droplet formation in the star. 
$N$ is uncertain,
but is at most the total baryon number 
in the star (typically $\sim 10^{57}$). 

Finally, we give some astrophysical implications. 
The nuclear equation of state of intermediate stiffness
used here allows a neutron star to have the central 
density higher than the transition point $\simeq 5n_{0}$.$^{12)}$ 
Once the central pressure reaches the pressure giving
a realistic value of $\tau/N$
during the spin-down or accretion, a droplet of $u$ and $d$ quarks may 
appear in the nuclear matter. We can see from Fig.\ 3 that
the electron screening, without which $E_{C}$ $(\propto R^{5})$ dominates
$U(R)$ for large $R$,
prevents the droplet from remaining finite for $R\lsim10$ fm.
For $R\gg\kappa^{-1}$, $E_{C}$ behaves as $E_{C}\propto R^{2}$
and $E_{C}\lsim 4\pi\sigma R^{2}$, 
hence $U(R)<0$.
The droplet may thus expand into bulk matter 
with the speed of sound, induced by the 
density jump due to deconfinement. 
The resultant bulk quark matter occupying the region $P\gsim P_{0}$ 
may change into three-flavor 
quark matter via weak interactions 
in a time scale of $10^{-9}$ sec.$^{13)}$ As long as three-flavor quark 
matter is more favorable than nuclear matter, 
the region of $u, d$, and $s$ quarks may spread$^{13)}$
with a release of chemical energy which amounts to 
$\sim100$ MeV per baryon as estimated from Fig.\ 1(a) and
turns immediately into thermal energy.
This implies that the mixed phase consisting of quark and nuclear 
matter$^{1),2)}$
is unlikely to occur. The resulting star could be stable,$^{14)}$ so
the thermal energy deposited in the star $\sim10^{52}$ ergs might 
be a possible origin of a $\gamma$-ray burst.$^{15)}$

In summary, we have found that a droplet of $u$ and $d$ quarks, if it
appeared in a neutron star, would develop into bulk matter due to
the electron screening effects. The effect of muons, which was
neglected above, makes 
nuclear matter more stable ($P_{0}\approx465$ 
MeV fm$^{-3}$) and 
shortens the TF screening length by $\simeq 1$ fm. We have confirmed that
such a change makes no important difference. 
In order to make better estimates, we should examine
dissipation effects on the nucleation, which may depend 
on the superfluidity of nuclear matter
(e.g., Ref.\ 16))
and control the crossover temperature from classical$^{17)}$ to 
quantum nucleation, 
the roles of hyperons
in the composition of matter before and after deconfinement, and
the poorly-known interfacial properties,
the effects of curvature from the quark side 
being shown to significantly destabilize the droplet.$^{18)}$
In addition to
recent low-temperature experiments,$^{19)}$
astrophysical dense matter such as neutron star matter of 
temperature $\lsim 0.1$ MeV may give us 
examples of quantum nucleation at first-order phase transitions.

\vspace{0.1cm}
This work was supported in part by
Grants-in-Aid for Scientific Research provided by the Ministry of
Education, Science, and Culture of Japan through Research 
Grants Nos. 05243103, 07CE2002, and 4396.

\end{document}